\documentclass[aps,prl,twocolumn,showpacs,superscriptaddress]{revtex4}
\usepackage{graphicx}
\usepackage{bm}
\usepackage{color}

\begin{document}

\title{Unified Phase Diagram for Iron-Based Superconductors}

\author{Yanhong Gu}
\affiliation{Beijing National Laboratory for Condensed Matter Physics, Institute of Physics, Chinese Academy of Sciences, Beijing 100190, China}
\affiliation{School of Physical Sciences, University of Chinese Academy of Sciences, Beijing 100190, China}
\author{Zhaoyu Liu}
\affiliation{Beijing National Laboratory for Condensed Matter Physics, Institute of Physics, Chinese Academy of Sciences, Beijing 100190, China}
\affiliation{School of Physical Sciences, University of Chinese Academy of Sciences, Beijing 100190, China}
\author{Tao Xie}
\affiliation{Beijing National Laboratory for Condensed Matter Physics, Institute of Physics, Chinese Academy of Sciences, Beijing 100190, China}
\affiliation{School of Physical Sciences, University of Chinese Academy of Sciences, Beijing 100190, China}
\author{Wenliang Zhang}
\affiliation{Beijing National Laboratory for Condensed Matter Physics, Institute of Physics, Chinese Academy of Sciences, Beijing 100190, China}
\affiliation{School of Physical Sciences, University of Chinese Academy of Sciences, Beijing 100190, China}
\author{Dongliang Gong}
\affiliation{Beijing National Laboratory for Condensed Matter Physics, Institute of Physics, Chinese Academy of Sciences, Beijing 100190, China}
\affiliation{School of Physical Sciences, University of Chinese Academy of Sciences, Beijing 100190, China}
\author{Ding Hu}
\affiliation{Beijing National Laboratory for Condensed Matter Physics, Institute of Physics, Chinese Academy of Sciences, Beijing 100190, China}
\author{Xiaoyan Ma}
\affiliation{Beijing National Laboratory for Condensed Matter Physics, Institute of Physics, Chinese Academy of Sciences, Beijing 100190, China}
\affiliation{School of Physical Sciences, University of Chinese Academy of Sciences, Beijing 100190, China}
\author{Chunhong Li}
\affiliation{Beijing National Laboratory for Condensed Matter Physics, Institute of Physics, Chinese Academy of Sciences, Beijing 100190, China}
\author{Lingxiao Zhao}
\affiliation{Beijing National Laboratory for Condensed Matter Physics, Institute of Physics, Chinese Academy of Sciences, Beijing 100190, China}
\affiliation{School of Physical Sciences, University of Chinese Academy of Sciences, Beijing 100190, China}
\author{Lifang Lin}
\affiliation{Beijing Normal University, Beijing 100875, China}
\author{Zhuang Xu}
\affiliation{Beijing Normal University, Beijing 100875, China}
\author{Guotai Tan}
\affiliation{Beijing Normal University, Beijing 100875, China}
\author{Genfu Chen}
\affiliation{Beijing National Laboratory for Condensed Matter Physics, Institute of Physics, Chinese Academy of Sciences, Beijing 100190, China}
\affiliation{School of Physical Sciences, University of Chinese Academy of Sciences, Beijing 100190, China}
\affiliation{Collaborative Innovation Center of Quantum Matter, Beijing 100190, China}
\author{Zi Yang Meng}
\affiliation{Beijing National Laboratory for Condensed Matter Physics, Institute of Physics, Chinese Academy of Sciences, Beijing 100190, China}
\author{Yi-feng Yang}
\affiliation{Beijing National Laboratory for Condensed Matter Physics, Institute of Physics, Chinese Academy of Sciences, Beijing 100190, China}
\affiliation{School of Physical Sciences, University of Chinese Academy of Sciences, Beijing 100190, China}
\affiliation{Collaborative Innovation Center of Quantum Matter, Beijing 100190, China}
\author{Huiqian Luo}
\email{hqluo@iphy.ac.cn}
\affiliation{Beijing National Laboratory for Condensed Matter Physics, Institute of Physics, Chinese Academy of Sciences, Beijing 100190, China}
\author{Shiliang Li}
\email{slli@iphy.ac.cn}
\affiliation{Beijing National Laboratory for Condensed Matter Physics, Institute of Physics, Chinese Academy of Sciences, Beijing 100190, China}
\affiliation{School of Physical Sciences, University of Chinese Academy of Sciences, Beijing 100190, China}
\affiliation{Collaborative Innovation Center of Quantum Matter, Beijing 100190, China}

\begin{abstract}
High-temperature superconductivity is closely adjacent to a long-range antiferromagnet, which is called a parent compound. In cuprates, all parent compounds are alike and carrier doping leads to superconductivity, so a unified phase diagram can be drawn. However, the properties of parent compounds for iron-based superconductors show significant diversity and both carrier and isovalent dopings can cause superconductivity, which casts doubt on the idea that there exists a unified phase diagram for them. Here we show that the ordered moments in a variety of iron pnictides are inversely proportional to the effective Curie constants of their nematic susceptibility. This unexpected scaling behavior suggests that the magnetic ground states of iron pnictides can be achieved by tuning the strength of nematic fluctuations. Therefore, a unified phase diagram can be established where superconductivity emerges from a hypothetical parent compound with a large ordered moment but weak nematic fluctuations, which suggests that iron-based superconductors are strongly correlated electron systems.

\end{abstract}



\maketitle

Iron pnictides share some common behaviors with many other unconventional superconductors, such as cuprates and some heavy-fermion superconductors, where superconductivity is achieved by suppressing the long-range antiferromagnetic (AFM) order in parent compounds \cite{DavisJCS13,ScalapinoDJ12}. What we learn from cuprates is that all parent compounds can be treated as Mott insulators and a unified phase diagram can thus be drawn \cite{LeePA06}. Superconductivity can be obtained by either hole or electron doping, suggesting that carrier doping can be directly associated with a microscopic quantum parameter. These consensus in cuprates naturally leads to the use of similar terms in iron-based superconductors, such as parent compound, electron and hole doping \cite{DaiP15}, despite the fact that there are many phenomena querying these simple adaptions. For example, the ordered moments of the AFM ground states in the parent compounds of iron pnictides vary significantly \cite{CruzC08,ZhaoJ08,ZhaoJ08b,ZhaoJ08c,ChenY08,HuangQ08,LiS09a}, leading to many theoretical efforts, but a consensus has not been reached yet \cite{YinZP11,YinZP11b,ToschiA12,TamYT15}. In NaFe$_{1-x}$Co$_x$As \cite{WangAF12}, filamentary superconductivity can be found in the AFM parent compound. The differences among these materials seem to disqualify them as the parent compound, which has long thought to be a Mott insulator \cite{SiQ08,FangC08,XuC08,IshidaH10,YuR13}. Efforts to find such an insulating parent compound have not met with much success \cite{DagottoE13,FangMH11,YanYJ11}. Moreover, achieving superconductivity in iron-based superconductors can be done by not just carrier doping but also isovalent doping \cite{KasaharaS10}.  It has been found that chemical substitution leads to the reduction of electronic correlations in most systems \cite{YeZR14}, but the reason is unknown. These diversities for both parent compounds and their doped materials make it hard to obtain a general picture for the low-energy physics of iron pnictides as that in cuprates, which seems to suggest that our understanding of antiferromagnetism and superconductivity in iron-based superconductors is not generic but material dependent.

A unique feature for the AFM order in iron pnictides is that it is always closely accompanied by a nematic order, which breaks the in-plane $C_4$ rotational symmetry of the high-temperature tetragonal lattice structure while preserving the translational symmetry \cite{ChuJH10,FernandesRM14}. Similar nematic order has also been found in cuprates but it is rather associated with the pseudogap \cite{DaouR10}. It has been shown that the nematic order and its fluctuations may overshadow the whole phase diagram of iron-based superconductors \cite{ChuJH12,KuoHH16,HosoiS16,ThorsmolleVK16,LiuZ16}. Moreover, a nematic quantum critical point (QCP) has been found in many near optimally doped iron pnictides \cite{KuoHH16,HosoiS16,ThorsmolleVK16,LiuZ16}. It has been theoretically suggested that quantum nematic fluctuations may induce attractive pairing interaction and thus enhance or even lead to superconductivity \cite{KimYB04,LedererS15}. Therefore, nematicity should play a significant role in both antiferromagnetism and superconductivity, and a quantitative relationship between the AFM and nematic orders may provide key information on the unified phase diagram of iron-based superconductors.

Here, we perform a detailed investigation on the nematic susceptibility of a variety of iron-based superconductors by studying the uniaxial pressure effect on the in-plane resistivity. The temperature dependence of the nematic susceptibility above the structural or nematic transition temperature $T_s$ can be well fitted by a Curie-Weiss-like function, which gives a nematic Curie constant $A_n$. We show that a linear relationship exists between $\left|A_n\right|^{-1}$ and the ordered moments in several classes of parent compounds and doped samples. Our results suggest that the suppression of the AFM order and the emergence of superconductivity are associated with the enhancement of nematic fluctuations. Effectively, these so-called parent compounds and their underdoped samples may all lie on a line with a general parent compound at one end and a QCP at the other. This picture unifies various iron-based superconductors into a coherent phase diagram.

\emph{R}FeAsO, NaFe$_{1-x}$Ni$_x$As, Ca$_{1-x}$La$_x$FeAs$_2$ and Ba$_{1-x}$K$_{x}$Fe$_2$As$_2$ single crystals were grown by flux methods similar to those reported previously \cite{YanJQ09,TanG13,XieT17,WangM13}. FeSe single crystals were grown through chemical vapor transport (CVT) method using AlCl$_3$ as the transport agent. Polycrystalline FeSe was first synthesized as predecessor using a solid state reaction by reacting stoichiometric amounts of high purity Fe and Se powders. The predecessor was reground and sealed in an evacuated quartz tube with AlCl$_3$ (3 mg/cm3), which was loaded into a horizontal two-zone furnace with the source end at 400 $^{\circ}$C and the sink end at 300 $^{\circ}$C. Flakelike single crystals were obtained at the cold end after a few months.
The uniaxial pressure dependence of resistivity is measured by the device reported previously \cite{LiuZ16}. The resistance shows a linear dependence on pressure above $T_s$ for all the samples measured here. We define $\zeta_{(110)}$ = $d(\Delta R/R_0)/dp$, where $R_0$ is the resistance at zero pressure and $\Delta R$ = $R(p)-R_0$ is the change of resistance under pressure $p$ along the tetragonal (110) direction. As shown in our previous study \cite{LiuZ16}, $\zeta_{(110)}$ can be well fitted by $A/(T-T')+y_0$, where $A$, $T'$, and $y_0$ are all temperature-independent parameters. Here $T'$ is always a few Kelvins lower than $T_s$ due to coupling between the nematic system and the lattice \cite{ChuJH12}. While the origin of $y_0$ is unknown, it should not be related to nematic fluctuations. We therefore obtain the nematic susceptibility $\chi_n$ as $\zeta_{(110)} - y_0$.

Figure 1 gives the temperature dependence of $\left|\chi_n\right|^{-1}$ in several classes of undoped and doped iron-based superconductors. The $T'$ value of FeSe is much larger than the similar Curie-Weiss temperature obtained by measuring the resistivity change with strain in Ref. \cite{HosoiS16} but close to that in Ref. \cite{TanatarMA16}. For near optimally doped BaFe$_2$(As$_{0.69}$P$_{0.31}$)$_2$, Ba$_{0.67}$K$_{0.33}$Fe$_2$As$_2$, and NaFe$_{0.985}$Ni$_{0.015}$As, $T'$ is close to zero, which is consistent with the presence of nematic QCPs \cite{KuoHH16}. It should be noted that the measurements have been done several times for materials with only small single crystals available in order to calculate the mean value of $A$ in the above fitting and its standard deviation.

Since $A$ is obtained from resistivity measurement, the effect of Fermi surfaces has to be considered. We introduce a dimensionless parameter $\kappa$ as $(\left|\nu_F^{\Gamma}\right|+\left|\nu_F^M\right|)^2$ normalized by that in BaFe$_2$As$_2$, where $\nu_F^{\Gamma}$ and $\nu_F^M$ are the Fermi velocities at the respective hole and electron pockets above $T_s$ that are nearly nested. Such a form is associated with the scattering of spin fluctuations in the dirty limit \cite{FernandesRM11}, where the resistivity anisotropy is proportional to the square of the difference of the $\Phi$ functions between electron and hole pockets. By definition, the $\Phi$ function is the projection of the Fermi velocity along the electric field direction. However, given the approximation and complication of the transport theory, we would like to emphasize that $\kappa$ was introduced as a phenomenological parameter rather than an established theoretical fact.

\begin{figure}
\includegraphics[scale=0.35]{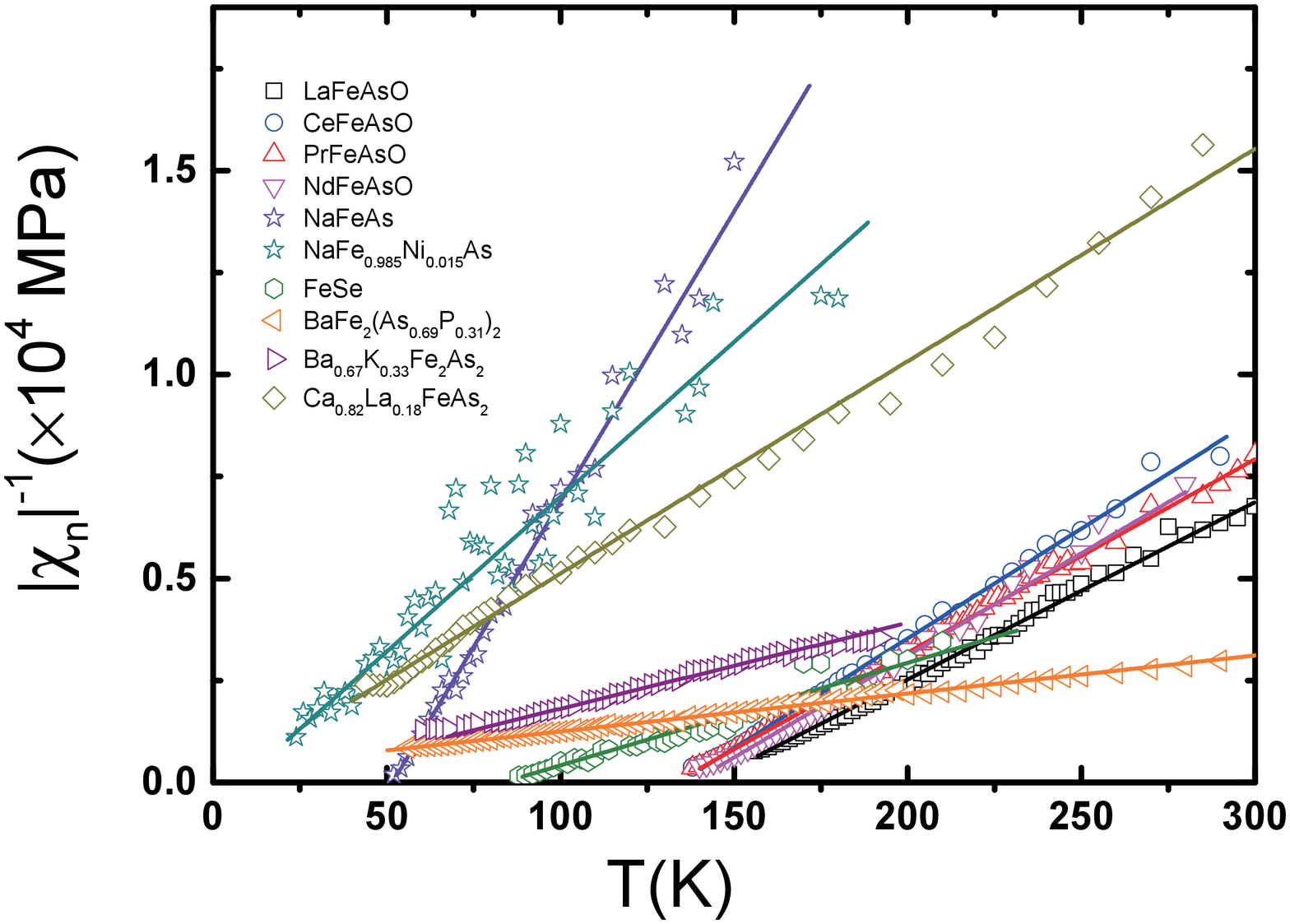}
\caption{Temperature dependence of $\left|\chi_n\right|^{-1}$ for various iron-based superconductors. The data include \emph{R}FeAsO (\emph{R} = La, Ce, Pr, Nd), NaFe$_{1-x}$Ni$_x$As ($x$=0, 0.015), FeSe, BaFe$_2$(As$_{0.69}$P$_{0.31}$)$_2$, Ba$_{0.67}$K$_{0.33}$Fe$_2$As$_2$, and Ca$_{0.82}$La$_{0.18}$FeAs$_2$. Only one set of data is shown for each kind of materials. For FeSe, NaFe$_{0.985}$Ni$_{0.015}$As, and Ba$_{0.67}$K$_{0.33}$Fe$_2$As$_2$, the values of $\chi_n$ are negative. The solid lines are linear fitting results.}
\label{fig1}
\end{figure}

Figure 2(a) gives the main result of this Letter, showing that the size of the ordered moment $M$ manifestatively scales with $\left|A_n\right|^{-1}$ in a linear fashion as long as it is not zero, where $A_n$ is defined as $\kappa A$ \cite{note}. The results include the parent compounds of three major classes of iron-based superconductors, i.e., ``1111" (\emph{R}FeAsO,where \emph{R} = La, Ce, Pr, Nd), ``Ba-122" (BaFe$_2$As$_2$), and ``111" (NaFeAs). Figure 2(a) also includes the results of BaFe$_{2-x}$Ni$_x$As$_2$ \cite{LiuZ16} that show the same behavior as the above parent compounds, although Ni dopants are supposed to provide electrons into the system and the doped samples should not be treated as parent compounds. The result of Ca$_{0.82}$La$_{0.18}$FeAs$_2$ ( ``112" ), which shows filamentary superconductivity \cite{XieT17}, also falls onto the same line. In the end, the data points for four samples without the AFM order, i.e., near optimally doped NaFe$_{1.985}$Ni$_{0.015}$As, BaFe$_2$(As$_{0.69}$P$_{0.31}$)$_2$, Ba$_{0.67}$K$_{0.33}$Fe$_2$As$_2$, and FeSe, are all at the right side of the data of optimally doped BaFe$_{2-x}$Ni$_x$As$_2$. A special point is that of SrFe$_2$As$_2$,  which is far away from the materials mentioned above. This is most likely due to the strong first-order AFM transition as discussed later. It should be noted that while the values of the Fermi velocities in calculating $\kappa$ are obtained from the measurements of the angle-resolved photoemission spectroscopy \cite{note}, we find that they give consistent results in our analysis. The most significant effects of $\kappa$ occur in NaFe$_{1-x}$Ni$_x$As ($\kappa \approx$ 11) and FeSe ($\kappa \approx$ 4), where the uncertainties will not significantly affect the positions of these two systems in Fig. 2(a) since their $\left|A_n\right|^{-1}$ is already very small.

The Curie-Weiss temperature dependence of the magnetic susceptibility gives a Curie constant $C$, which is associated with magnetic fluctuations. Since the nematic order in iron-based superconductors is an Ising-type order \cite{FernandesRM14}, we call $A_n$ as the nematic Curie constant. Similarly, $A_n$ is associated with nematic fluctuations, which come from fluctuations of local effective nematic moment. The value of local nematic moment may not be equal to that of the long-range ordered nematic moment if quantum fluctuations present \cite{LiuZ16}, although it seems that the resistivity anisotropy at zero temperature becomes larger with increasing Co doping in Ba(Fe$_{1-x}$Co$_x$)$_2$As$_2$ \cite{ChuJH12}.

Figure 2(a) is the first result that connects the size of the ordered magnetic moment to another experimental data. It shows that the ground state of antiferromagnetism is associated with high-temperature nematic fluctuations, which is totally unexpected especially considering that the system goes through a nematic transition first with decreasing temperature before entering the long-range AFM order. To reduce the ordered moment, theoretical attempts have introduced frustrations into the magnetic system, such as temporal fluctuations and the nesting effect \cite{YinZP11,YinZP11b,ToschiA12,TamYT15}. Our results suggest that the same mechanism also enhances nematic fluctuations and thus put strong constraints on the microscopic origin of those frustrations. Naively, the strong nematic fluctuations suggest that either the direction or the amplitude of local nematic order fluctuates rapidly with time, which may result in a small value of ordered moment since the AFM order is linearly coupled to the nematic order \cite{FernandesRM14}. In other words, strong nematic fluctuations may frustrate the magnetic system and drive it away from establishing long-range AFM order.

\begin{figure}
\includegraphics[scale=0.85]{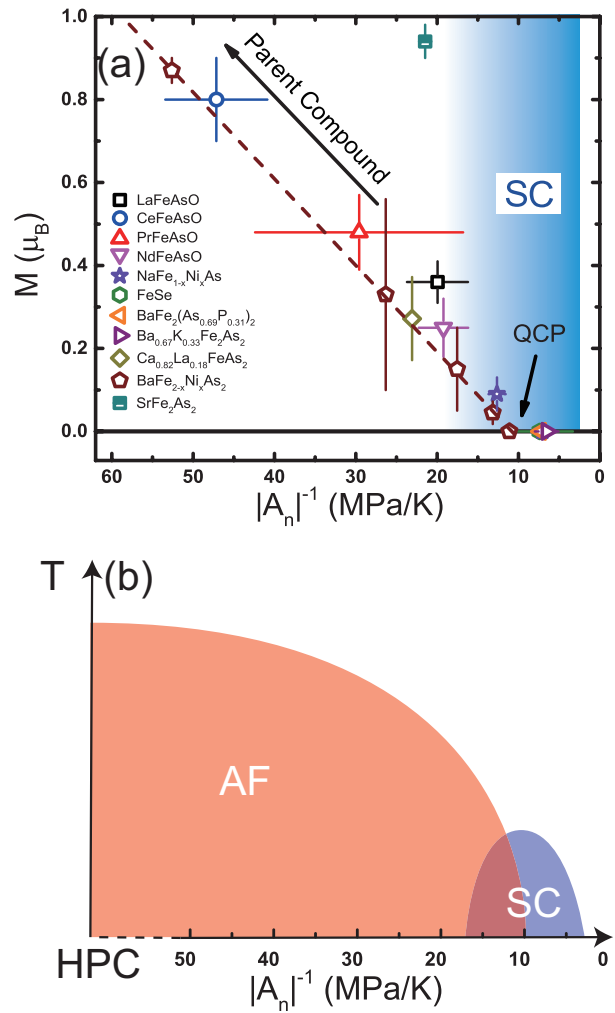}
\caption{(a) Scaling between ordered moment $M$ and $\left|A_n\right|^{-1}$. The straight dashed line in the front panel is guided to the eye according to the data of BaFe$_{2-x}$Ni$_x$As$_2$. Superconductivity (SC) tends to appear with sufficient small $\left|A_n\right|^{-1}$. The values and error bars of magnetic moments are from previous measurements by neutron diffraction \cite{CruzC08,ZhaoJ08,ZhaoJ08b,ZhaoJ08c,ChenY08,HuangQ08,LiS09a,XieT17,LuoH12}. The vertical error bars are the standard deviations of multiple measurements. The values of $A_n$ for BaFe$_{2-x}$Ni$_x$As$_2$ are obtained from previous works \cite{LiuZ16} assuming Fermi velocities change little with Ni doping. The data points of BaFe$_2$(As$_{0.69}$P$_{0.31}$)$_2$, Ba$_{0.67}$K$_{0.33}$Fe$_2$As$_2$, FeSe, and NaFe$_{1.985}$Ni$_{0.015}$As overlap each other. (b) Schematic phase diagram of iron-based superconductors where superconductivity (SC) is achieved by tuning $\left|A_n\right|^{-1}$ to suppress the AFM order in a hypothetical parent compound (HPC). It should be pointed out that the values of $T_N$ and $T_c$ are material dependent.}
\label{fig2}
\end{figure}

The roughly linear relationship in Fig. 2(a) can be further understood from the case of BaFe$_{2-x}$Ni$_x$As$_2$. Doping Ni into BaFe$_2$As$_2$ gradually suppresses the long-range AFM order and superconductivity appears above 5\% of the Ni doping level \cite{LuoH12}. The optimal doping level is achieved at about 0.1, where the AFM order also disappears \cite{LuoH12}. It is a typical system of iron-based superconductors that superconductivity is described as introduced by electron doping. The dashed line in Fig. 2(a) suggests that this process can also be described from another point of view, i.e., Ni doping reduces $\left|A_n\right|^{-1}$, which leads to the suppression of the long-range AFM order and the emergence of superconductivity.

While these two descriptions are indistinguishable in BaFe$_{2-x}$Ni$_x$As$_2$, the latter may be more reasonable since this relationship also holds for other compounds as shown in Fig. 2(a). For example, while \emph{R}FeAsO and NaFeAs are the so-called parent compounds, their magnetic ground states may be simply achieved by tuning the strength of $\left|A_n\right|^{-1}$. All the optimally doped materials listed in Fig. 2(a) have rather small $\left|A_n\right|^{-1}$, which suggests that the main role played by dopants can be understood as enhancing nematic fluctuations to suppress the AFM order. Interestingly, it seems that FeSe should show no magnetic order according to Fig. 2(a) because its $\left|A_n\right|^{-1}$ is very small. Whether the competition between various magnetic orders in this system \cite{LiW17} results in its small value of $\left|A_n\right|^{-1}$ needs to be further addressed.

The above discussions suggest that a unified phase diagram for iron pnictides may be established as shown in Fig. 2(b). Compared to the usual way of plotting the phase diagram, the most important difference is that the tuning parameter is not the carrier doping level but $\left|A_n\right|^{-1}$ (or the effective nematic moment). There are three key aspects in this phase diagram. First, there is a hypothetical parent compound (HPC) with a large AFM ordered moment and weak nematic fluctuations. The magnetic ground states of undoped materials are of course determined by their electronic configurations, but they can be effectively treated as coming from the same HPC by tuning the strength of $\left|A_n\right|^{-1}$. In this sense, none of the materials studied here is sufficiently qualified as a parent compound since they are rather the same as those doped materials if we just consider their magnetic ground states. The second aspect is that there may be a nematic quantum critical point (QCP), which is widely seen in many iron-based superconductors \cite{KuoHH16,HosoiS16,ThorsmolleVK16,LiuZ16}. Since the nematic QCP is close to the optimal doping level in many systems, it seems that the emergence of superconductivity is associated with the enhancement of $\left|A_n\right|$, which is the third aspect of the phase diagram. For example, it explains why NaFeAs as a parent compound can be easily doped to become superconductivity \cite{WangAF12} since its $\left|A_n\right|^{-1}$ is already small. Moreover, one needs not to distinguish electron, hole, and isovalent doping since superconductivity will appear as far as $\left|A_n\right|^{-1}$ is small enough. It may explain why FeSe is superconducting since its $\left|A_n\right|^{-1}$ is similar to those in optimally doped materials. It should be pointed out that the actual values of thermodynamic properties such as $T_N$ and $T_c$ should be material dependent. In addition, whether there is a coexisting region for nematic order and superconductivity may also vary from system to system \cite{LangG16}.

Our results support the idea of treating iron pnictides as strongly correlated systems \cite{DavisJCS13,ScalapinoDJ12,SiQ08,FangC08,XuC08,IshidaH10,YuR13,MaT13}, although whether the HPC is a Mott insulator or not remains unclear. Nevertheless, it is thus possible to establish a unified phase diagram of iron-based superconductors similar to cuprates but with the nonthermal parameter associated with nematic fluctuations instead of doping. 	Experimentally, the magnetic transition becomes strongly first order for materials with large moments. The sharp increase of $M$ just below $T_N$ also results in a jumplike behavior for nematic susceptibility \cite{GongD17}, which makes the above analysis inappropriate as shown by the data of SrFe$_2$As$_2$ in Fig. 2(a). In other words, the roughly linear relationship between $M$ and $A_n^{-1}$ is only valid for materials with the second-order or weakly first-order magnetic transition. Moreover, the spin structure changes with a further increasing ordered moment (e.g., FeTe \cite{BaoW09,LiS09b}). Therefore, the attempt to realize a true parent compound experimentally without changing the magnetic structure may be pointless.

In the end, we note that superconductivity has been found in many other iron-based materials \cite{HosonoH15}, where it is not clear whether the low-energy spin fluctuations are associated with the striped AFM order due to the lack of large-size single crystals. Moreover, a second superconducting dome has been found in a H-doped 1111 system \cite{HosonoH15} and more complicated superconducting and magnetic behaviors are discovered in FeSe under pressure \cite{SunJP16}. We have also limited our analysis within the systems that show superconductivity. Some dopants, e.g., Mn, Cr, and Cu, do not or hardly introduce superconductivity \cite{NiN10,TakedaH14,KobayashiT15}, which may be caused by factors other than nematic fluctuations, such as impurity scattering. Whether these phenomena can be understood within the interplay between magnetic and nematic correlations needs to be addressed in future investigations.

In conclusion, we find that the magnetic ground states of a variety of iron-based superconductors can be achieved by tuning the strength of nematic fluctuations, suggesting the dominant roles played by magnetic and nematic correlations. Accordingly, a unified phase diagram can be established where superconductivity comes from the suppression of the AFM order in a hypothetical parent compound by enhancing nematic fluctuations. Our results provide general understandings on some of the key phenomena in iron-based superconductors, such as the variety of ordered moments in the so-called parent compounds and the role of isovalent dopants in generating superconductivity.

\acknowledgements

S. L. thanks Professor Tao Xiang, Professor Jiangping Hu, Professor Xianggang Qiu, Professor Daoxin Yao, and Professor Wei Ku for helpful discussion. Y. G. and Z. L. are grateful for help from Jia Yu, and Wei Zhang. This work is supported by the Ministry of Science and Technology of China (No. 2017YFA0302903, No. 2016YFA0300502, and No. 2015CB921300), the National Natural Science Foundation of China (No. 11674406, No. 11374346, No. 11374011, No. 11305257, No. 11674372, No. 11421092, and No. 11574359, No. 11522435, No. 11774401), the Strategic Priority Research Program(B) of the Chinese Academy of Sciences (XDB07020300,XDB07020200, XDPB01), and the Fundamental Research Funds for the Central Universities (Grants No. 2014KJJCB27). H. L. and Y. Y. are supported by the Youth Innovation Promotion Association of CAS. Z. Y. M. is supported by the National Thousand-Young-Talents Program of China.

\end{document}